# High-temperature Superconductivity in Perovskite Hydride below 10 GPa

Mingyang Du[1], Hongyu Huang[1], Zihan Zhang[2], Min Wang[1], Hao Song[1,*], Defang Duan[2], Tian Cui[1,2,*]

[1] Institute of High Pressure Physics, School of Physical Science and Technology, Ningbo University, Ningbo, 315211, People's Republic of China

[2] College of Physics, Jilin University, Changchun 130012, People's Republic of China

**ABSTRACT**

Hydrogen and hydrides materials have long been considered promising materials for high-temperature superconductivity. But the extreme pressures required for the metallization of hydrogen-based superconductors limit their applications. Here, we have designed a series of high-temperature perovskite hydrides that can be stable within 10 GPa. Our research covered 182 ternary systems and ultimately determined that 9 compounds were stable within 20 GPa, of which 5 exhibited superconducting transition temperatures exceeding 120 K within 10 GPa. Excitingly, $KGaH_3$ and $CsInH_3$ are thermodynamically stable at 50 GPa. Among these perovskite hydrides, alkali metals are responsible for providing a fixed amount of charge and maintaining structural stability, while the cubic framework formed by IIIA group elements and hydrogen is crucial for high-temperature superconductivity. This work will inspire further experimental exploration and take an important step in the exploration of low-pressure stable high-temperature superconductors.

**Introduction**

The discovery of superconductors with high transition temperatures ($T_c$) has been a long-standing hot topic in the scientific community. Since the discovery of superconductivity of mercury in 1911, room-temperature superconductivity has been a long-sought dream and a field of intensive research. In 1935, E. Winger and H. B. Huntington theoretically predicted that solid metal hydrogen can be obtained under high pressure conditions[1]. According to the BCS theory, the superconducting transition temperature of material is proportional to its Debye temperature[2]. This suggests that hydrogen, the lightest element in nature, would be an ideal room-temperature superconductor after metallization[3]. However, experimental studies have shown that hydrogen requires extremely high pressures to metallize[4, 5]. As a result, the search for room-temperature superconductor has gradually turned to another, more feasible route — hydrogen-rich compounds.

By incorporating other elements to create a "chemical pre-compression" effect on hydrogen, hydrogen-rich compounds can metallize at much lower pressures than pure hydrogen[6]. Guided by this principle, many excellent hydrogen-rich compounds have been designed and predicted to be potential high-temperature superconductors in the past decade[7-9], some of which have been experimentally confirmed. In particular, $H_3S$ and $LaH_{10}$ were first predicted to have superconducting transition temperatures ($T_c$s) exceeding 200 K[10-12], which have been confirmed experimentally[13-16]. These are important milestones in the exploration of hydrogen-based superconductors.

Among these hydrogen-based superconductors, clathrate superhydrides have attracted extensive attention due to their outstanding superconductivity. They are widely found in alkaline earth metals and rare earth element superhydrides RH$_n$ (n = 6, 9, 10), such as binary hydrides CaH$_6$[17], MgH$_6$[18], YH$_{6,9,10}$[11, 12, 19], ScH$_6$[20], (Tm/Yb/Lu)H$_6$[21], and ternary hydrides (Y,Ca)H$_6$[22-24], (Mg,Ca)H$_6$[25], (Sc,Ca)H$_6$[26], (La,Y)H$_6$[27], (Ca/Sc/Y,Yb/Lu)H$_6$[28]. In the structure of these superhydrides, H atoms are weakly covalently bonded each other to form H$_{24}$, H$_{29}$ and H$_{32}$ cage, and the metal atom is located in the center of the H cage. After LaH$_{10}$ confirmed the great potential of clathrate superhydrides in high-temperature superconductivity, more clathrate superhydrides were experimentally synthesized, including ThH$_9$[29], ThH$_{10}$[29], YH$_9$[30-32], CaH$_6$[33, 34], YH$_6$[35], CeH$_9$[36], (La,Y)H$_{10}$[37], (La,Ce)H$_{9,10}$[38, 39].

In this work, we have designed a series of high-temperature superconductors that can be stable within 10 GPa based on perovskite structure of $Pm$-3$m$-KInH$_3$[40]. Our research covered 182 ternary systems and ultimately determined that 9 compounds were stable within 20 GPa, of which 5 exhibited superconducting transition temperatures exceeding 120 K within 10 GPa. Excitingly, KGaH$_3$ and CsInH$_3$ are thermodynamically stable at 50 GPa. Among these perovskite hydrides, alkali metals are responsible for providing a fixed amount of charge and maintaining structural stability, while the cubic framework formed by IIIA group elements and hydrogen is crucial for high-temperature superconductivity. This work will inspire further experimental exploration and take an important step in the exploration of low-pressure stable high-temperature superconductors.

**Computational details**

Ab initio random structure searching (AIRSS) technique[41, 42] and ab initio calculation of the Cambridge Serial Total Energy Package (CASTEP)[43] were used to predict the candidate crystal structures of AXH$_3$. The generalized gradient approximation with the Perdew-Burke-Ernzerhof parametrization[44] for the exchange-correlation functional and on-the-fly generation of ultra-soft potentials were used for the structure searching.

The Vienna ab initio simulation program (VASP)[45] was used for structural relaxation and calculations of enthalpies and electronic properties. The projector augmented plane-wave potentials[46] with an energy cutoff of 800 eV and Monkhorst-Pack[47] meshes for Brillouin zone sampling with resolutions of 2π×0.03 Å$^{-1}$ were used to ensure that all enthalpy calculations are well converged to less than 1 meV per atom.

The Quantum-ESPRESSO[48] was used in phonon and electron−phonon calculations. Ultra-soft potentials were used with a kinetic energy cut-off of 90 Ry. The k-points and q-points meshes in the first Brillouin zone are 12×12×12 and 6×6×6. The superconducting transition temperatures of these structures are estimated through the self-consistent iteration solution of the Eliashberg equation (scE)[49].

**Results and discussion**

Using the perovskite structure of *Pm*-3*m*-KInH$_3$ as a template, we obtained a series of compounds AXH$_3$ by substituting the IA-IVA elements and IIB elements, with the structure shown at the middle of Fig. 1. The A indicates Alkali metals, alkaline earth metals, rare earth metals, and the X indicates IIIA or IVA elements. All the elements involved in this study are shown in Fig. 1.

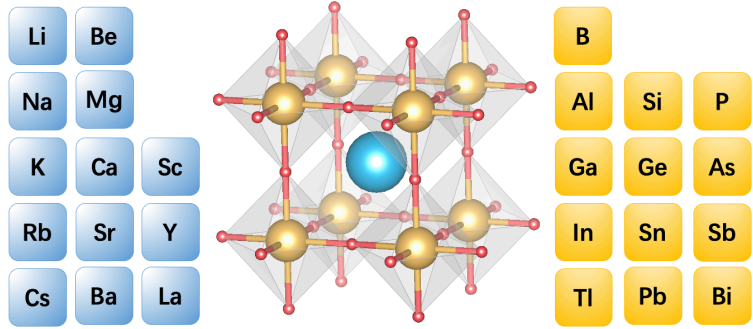

Fig. 1. Crystal structure of the perovskite hydrides $AXH_3$. Blue, yellow, and red balls represent A (group IA, IIA or IIIB metal), X (group IIIA, IVA or VB element), and H atoms, respectively.

We first did a screening for the dynamical stability of perovskite hydrides $AXH_3$ with space group $Pm$-$3m$ at 0-50 GPa. The Quantum-ESPRESSO code was used to initially verify the dynamical stability of these compounds. In this study, we have uncovered nine dynamically stable perovskite hydrides, $KAlH_3$, $RbAlH_3$, $KGaH_3$, $RbGaH_3$, $KInH_3$, $RbInH_3$, $CsInH_3$, $RbTlH_3$, $CsTlH_3$, which are marked by circles in Fig. 2. The gray cross indicates that this structure cannot be stable within 50 GPa. Perhaps these structures can remain stable under higher pressures, but it is unlikely that these perovskite hydrides have $T_c$ exceeding 200 K, at pressures exceeding 50 GPa, they do not have significant advantages compared to other high-temperature superconducting materials. Therefore, in this work, we will not study perovskite hydrides that can only be stable above 50 GPa.

For dynamically stable ternary perovskite hydrides, we further determined their thermodynamic stability. We built a database for high pressure phase of elements and binary hydrides at 0-50 GPa. Ab initio random structure searches were performed in each $AXH_3$ compound at pressure of 0-50 GPa to supplement the enthalpy information of ternary compounds. Based on the database, a high-throughput code was developed to construct ternary convex hull graphs. The enthalpy values deviating from the ternary convex hull line are shown as the color of markers in Fig. 2. The bluer the color means more thermodynamically stable. The circular marks indicates that these structures are metastable within 50 GPa, and the square marks indicates that these structures are thermodynamically stable. In this study, we have uncovered three thermodynamically stable perovskite hydrides, $KGaH_3$, $RbGaH_3$ and $CsInH_3$. It is worth noting that for the structure $KInH_3$, which was previously mentioned to be stable at ambient pressure, we have conducted calculations with various accuracies. Finally, it was found that our calculation result with q-point of 3×3×3 was in good agreement with previous work. But with q-point of 6×6×6, $KInH_3$ cannot stabilize at ambient pressure (see Fig. S2 in Supplemental Material). In our calculations, $KInH_3$ can only dynamically stabilize at pressure above 20 GPa.

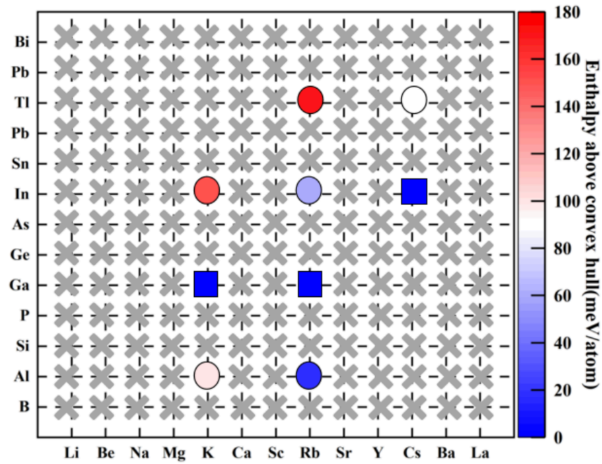

Fig. 2. The circular marks indicate the dynamically stable perovskite hydrides $AXH_3$ at 50 GPa. The square marks indicate the thermodynamically stable $AXH_3$ at pressure lower than 50 GPa. The gray cross represents the $AXH_3$ that are dynamically unstable at pressure lower than 50 GPa. The enthalpy values of $AXH_3$ above the convex hull line are represented by the color of markers. The color bar indicates the correspondence between the color and values.

We calculated the electron-phonon coupling (EPC) to estimate the superconductivity. The $T_c$s are estimated through the self-consistent iteration solution of the Eliashberg equation (scE) with $\mu^*$=0.10 as a function of pressure are supplied in Fig. 3. RbTlH$_3$ shows the highest $T_c$ of 170 K at 4 GPa, followed by CsTlH$_3$ with $T_c$ of 163 K at 7 GPa. But unfortunately, both of them are metastable phases, this means that some difficulties need to be overcome in the experiment to synthesize these structures. Considering the continuous progress in the synthesis technology of metastable materials in experiments, there is also a considerable possibility that we will see these excellent properties of metastable materials successfully synthesized in the future[50]. CsInH$_3$ shows the highest $T_c$ of 153 K at 9 GPa in all thermodynamically stable perovskite hydrides, followed by KGaH$_3$ with a $T_c$ of 146 K at 10 GPa. RbGaH$_3$ requires more than 20 GPa to be dynamically stable, and its $T_c$ is expected to be 127 K at this time. In addition, the metastable phase RbInH$_3$ also exhibits good properties, with a $T_c$ of 130 K at 6 GPa. Other metastable phase KAlH$_3$ and RbAlH$_3$ shows the $T_c$ of 95 K at 7 GPa and 86 K at 15 GPa respectively. Detailed computational results at different pressures are supplied in Table S1 in the Supplemental Material. In addition, the $T_c$s of these perovskite hydrides are more sensitive to pressure than other hydrides. RbTlH$_3$ shows the $T_c$ of 170 K at 4 GPa, but as the pressure increased to 20 GPa, its $T_c$ rapidly decreased to 81 K. The $T_c$ of KGaH$_3$, CsInH$_3$, and CsTlH$_3$ also rapidly decrease with increasing pressure.

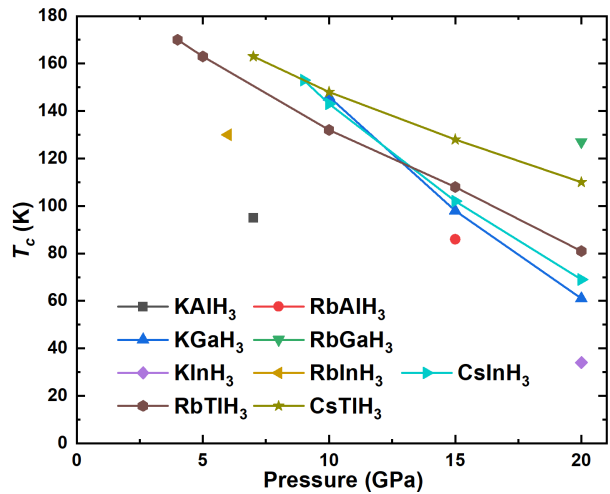

Fig. 3. The $T_c$s of perovskite hydrides AXH$_3$ estimated through the self-consistent iteration solution of the Eliashberg equation as a function of pressures.

We furthermore calculated the electron localization function (ELF) and Bader charge of these perovskite hydrides at different pressures (see Fig. 4). The structure of perovskite hydride AXH$_3$ can be divided into two parts: alkali metals, and the framework structure formed by IIIA group elements and hydrogen. Fig. 4a shows the ELF of the alkali metal layer of RbAlH$_3$, charge localization around alkali metal Rb. This indicates that the alkali metal Rb plays a "pre-compressor" role to support the framework structure formed by IIIA group elements and hydrogen in RbAlH$_3$ and does not directly contribute to superconductivity. The ELF of the alkali metal layer in other AXH$_3$ structures is basically consistent with Fig. 4a (see Fig. S3 in the Supplemental Material), which means that alkali metals play a "pre-compressor" role in all perovskite hydrides AXH$_3$. From Fig. 4e, it can be seen that the electrons provided by the alkali metal Rb in RbXH$_3$ (X = Al, Ga, In, Tl) remain in a relatively stable range (0.69 e - 0.77 e). Even if replaced with other alkali metals with different radii, the electrons provided by alkali metals in AInH$_3$ (A = K, Rb, Cs) remain within a relatively stable range (0.71 e - 0.72 e). This means that alkali metals also act as electron donors in AXH$_3$, and alkali metals provide a relatively fixed number of electrons. Changes in elements and pressure have little effect on the number of electrons provided by alkali metals.

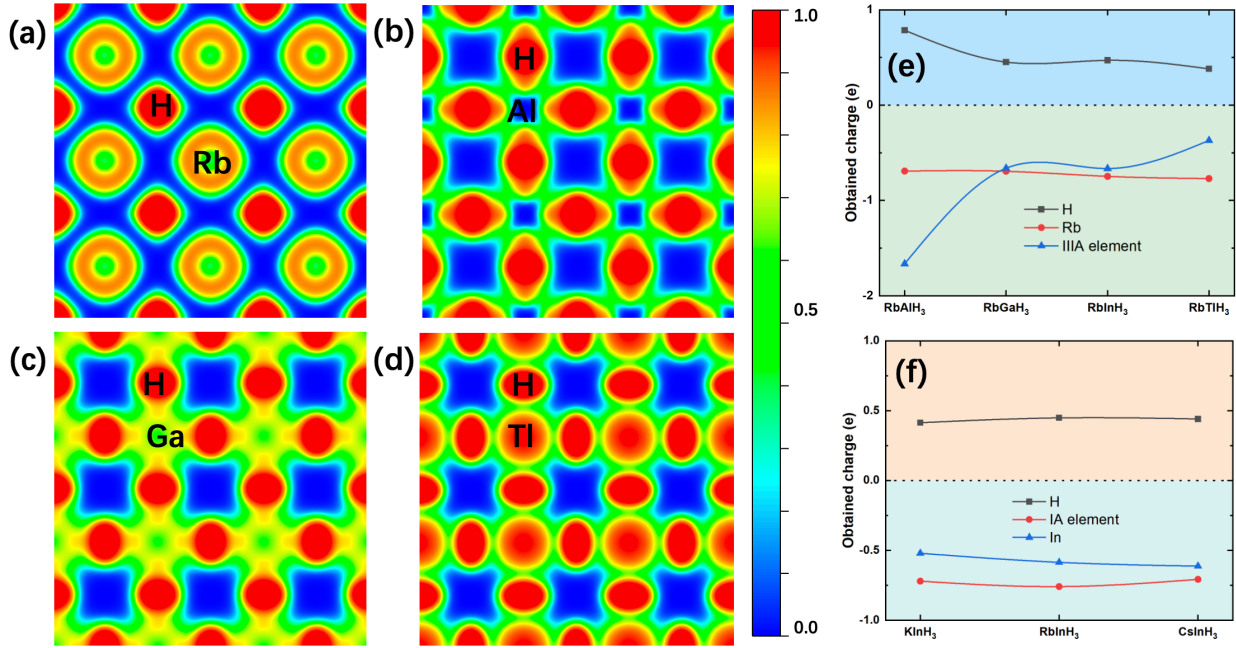

Fig. 4. The 2-dimensional electron localization function (ELF) for perovskite hydrides (a) $RbAlH_3$ at alkali metal layer, (b) $RbAlH_3$, (c) $RbGaH_3$ and (d) $RbTlH_3$ at IIIA group atomic layer. Remnant charges on A, X and H atoms obtained from Bader Charge Analysis of (e) $RbXH_3$ (X = Al, Ga, In, Tl) and (f) $AInH_3$ (A = K, Rb, Cs).

Fig. 4b-d shows the ELF of the IIIA group atomic layers in $RbAlH_3$, $RbGaH_3$ and $RbTlH_3$, respectively. The IIIA group atomic layers in different $AXH_3$ exhibit significant differences, both in charge density and charge distribution. The Al layer in $RbAlH_3$ has the lowest charge density in $RbXH_3$ (X = Al, Ga, In, Tl), and the charges around H are distributed in an ellipsoidal shape, with the ellipsoidal axis along the Al-H direction (see Fig. 4b). The Ga layer in $RbGaH_3$ has the highest charge density, and the charges around H are distributed in a spherical shape (see Fig. 4c), and the ELF of the In layer in $RbInH_3$ is also basically similar (see Fig. S3 in the Supplemental Material). The Tl layer in $RbTlH_3$ has the moderate charge density, and the charges around H are distributed in an ellipsoidal shape, with the ellipsoidal axis perpendicular to the Tl-H direction (see Fig. 4d). It can be seen that the X-H framework formed by IIIA group elements and hydrogen is the key to the superconductivity of the $AXH_3$ system, and the charge density and distribution on the B-H framework will affect the superconductivity of the system. From Fig. 4e, it can also be seen that there are significant differences in charge transfer between different IIIA group elements, with Al transferring up to 1.66 e of charge, while Tl only transferring 0.37 e of charge. Considering the superconductivity and stability of each $AXH_3$ in Fig.2-3, we believe that the Tl element with the least charge transfer (0.37 e) is more conducive to high-temperature superconductivity and low-pressure dynamic stability of the structure, while the Ga and In elements with moderate charge transfer (0.66 e) are more conducive to thermodynamic stability of the structure.

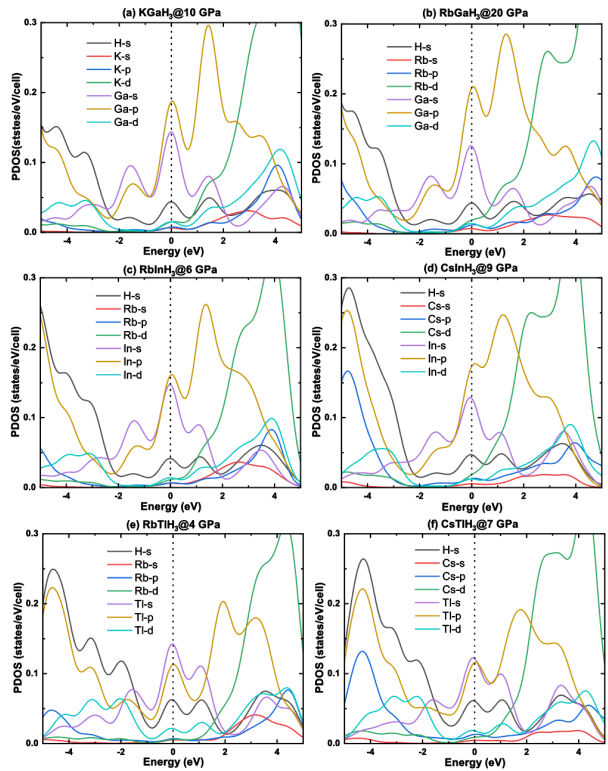

Fig. 5. The projected electronic density of states of (a) $KGaH_3$ at 10 GPa, (b) $RbGaH_3$ at 20 GPa, (c) $RbInH_3$ at 6 GPa, (d) $CsInH_3$ at 9 GPa, (e) $RbTlH_3$ at 4 GPa and (f) $CsTlH_3$ at 7 GPa.

Subsequently, we calculated the electronic band structures (see Fig. S4 in the Supplemental Material) and projected density of states (PDOS) to understand the effect of pressure on the electronic structure for KGaH$_3$ at 10 GPa, RbGaH$_3$ at 20 GPa, RbInH$_3$ at 3 GPa, CsInH$_3$ at 9 GPa, RbTlH$_3$ at 4 GPa and CsTlH$_3$ at 7 GPa. As shown in Fig. 5, AXH$_3$ share the similar electronic structure with peak of DOS located above the Fermi energy level (Ef), contributions from s-orbitals and p-orbitals in IIIA element dominate the DOS at the Fermi level. When the IIIA elements are the same, the electronic structure of AXH$_3$ is basically the same, which is also consistent with the results in Fig. 4f. The key difference in electronic structure of AXH$_3$ lies in the contribution of hydrogen on the Fermi surface. The contribution of H in the Fermi surface is higher in RbTlH$_3$ and CsTlH$_3$, which may be an important reason for their higher $T_c$. Based on the charge transfer situation in Fig. 4e, we believe that in AXH$_3$, less charge transfer is more favorable for the DOS peak of hydrogen near the Fermi surface, leading to a higher superconducting transition temperature. The X-H covalent framework is the key to AXH$_3$ high-temperature superconductivity. As hydrogen gains too many electrons and exhibits hydrogen ion properties, its DOS will appear more in deep energy levels rather than near the Fermi surface.

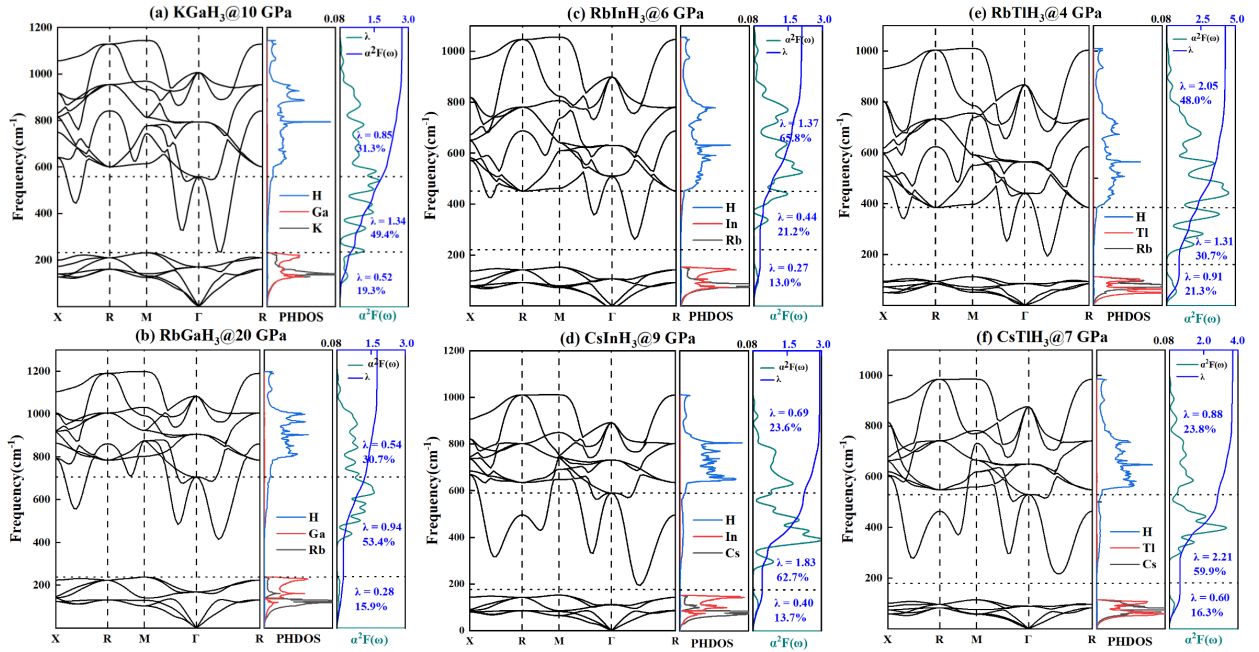

Fig. 6. Calculated phonon dispersion curves, phonon density of states (PHDOS), Eliashberg spectral function α$^2$F(ω) and the accumulated EPC constant λ of (a) KGaH$_3$ at 10 GPa, (b)

RbGaH$_3$ at 20 GPa, (c) RbInH$_3$ at 6 GPa, (d) CsInH$_3$ at 9 GPa, (e) RbTlH$_3$ at 4 GPa and (f) CsTlH$_3$ at 7 GPa.

To further understand the source of high temperature superconductivity of these perovskite hydrides AXH$_3$, we calculated the phonon band structure, projected phonon density of states (PHDOS), accumulated electron-phonon coupling (EPC) parameter λ, and Eliashberg spectral function $\alpha^2F(\omega)$ of KGaH$_3$ at 10 GPa, RbGaH$_3$ at 20 GPa, RbInH$_3$ at 3 GPa, CsInH$_3$ at 9 GPa, RbTlH$_3$ at 4 GPa and CsTlH$_3$ at 7 GPa, as shown in Fig. 6. It can be seen from PHDOS that the low frequency phonon modes (below 200 cm$^{-1}$) are mainly from IA and IIIA elements, high frequency phonon modes (above 200 cm$^{-1}$) are mainly from H atom. The accumulated EPC constant λ in the low-frequency region accounts for 13% - 21.3% of the total λ, indicating that non-hydrogen elements contribute less to the electron-phonon coupling. The key of these perovskite hydrides AXH$_3$ to high-temperature superconductivity, like other hydrides, is also related to the high-frequency vibration of hydrogen.

As we discussed in Fig. 3, the $T_c$s of these perovskite hydrides are more sensitive to pressure than other hydrides. From the phonon band structure, it is likely due to a significant contribution of phonon softening in the superconductivity of these structures. The accumulated EPC constant λ in the phonon softening region accounts for 21.2 % - 62.7% of the total λ, which is higher than the contribution of non-hydrogen elements. Furthermore, it should be noted that a considerable portion of the electroacoustic coupling in the high-frequency region also comes from phonon softening, such as phonon softening near the frequency of 600 cm$^{-1}$ on the high symmetry path X-R-M in RbInH$_3$ and RbTlH$_3$ (see Fig. 6c and 6e). And The growth of accumulated EPC constant λ is most rapid around the frequency around 600 cm$^{-1}$. Considering this, we can consider that phonon softening contributes more than 50% to the electron-phonon coupling among the six perovskite hydrides $AXH_3$ in Fig. 6. When the pressure increases, phonon softening weakens, and the corresponding electron-phonon coupling rapidly weakens, ultimately leading to the rapid decrease in $T_c$. The calculated EPC constant λ, logarithmic average phonon frequency $\omega_{log}$, superconducting critical temperature $T_c$ for perovskite hydrides $AXH_3$ at different pressures are list in Table. S3 in the Supplemental Material. It can be clearly seen that as the pressure increases to 20 GPa, the λ of $CsInH_3$ rapidly decreases from 2.92 to 1.24, and $T_c$ of $CsInH_3$ decreases from 153 K to 69 K; the λ of $RbTlH_3$ decreases from 4.27 to 1.28, and $T_c$ of $RbTlH_3$ decreases from 170 K to 81 K; the λ of $CsTlH_3$ decreases from 3.69 to 1.48, and $T_c$ of $RbTlH_3$ decreases from 163 K to 110 K. This high sensitivity of $T_c$ to pressure may result in higher requirements for precise pressure control in future experimental synthesis of such hydrides. But from another perspective, this sensitivity may allow them to be applied in more fields beyond high-temperature superconductivity.

**Conclusions**

In summary, we have designed a series of high-temperature superconductors that can be stable within 10 GPa based on perovskite structure of $Pm$-$3m$-$KInH_3$. Our research covered 182 ternary systems and ultimately determined that 9 compounds were stable within 20 GPa, of which 5 exhibited superconducting transition temperatures exceeding 120 K within 10 GPa, including $KGaH_3$ (146 K at 10 GPa), $RbInH_3$ (130 K at 6 GPa), $CsInH_3$ (153 K at 9 GPa), $RbTlH_3$ (170 K at 4 GPa) and $CsTlH_3$ (163 K at 7 GPa). Excitingly, $KGaH_3$ and $CsInH_3$ are thermodynamically stable at 50 GPa. Among these perovskite hydrides, alkali metals are responsible for providing a fixed amount of charge and maintaining structural stability, while the cubic framework formed by IIIA group elements and hydrogen is crucial for high-temperature superconductivity. EPC calculations revealed that the contribution of the softening phonon mode of hydrogen to the EPC constant λ exceeds 50%. This phonon softening enables these hydrides to exhibit unexpectedly high superconducting transition temperatures at low pressures, but as the pressure increases, the softening decreases, and their superconducting transition temperature rapidly decreases to less than half of its maximum $T_c$. This high sensitivity of $T_c$ to pressure may result in higher requirements for precise pressure control in future experimental synthesis of such hydrides. But from another perspective, this sensitivity may allow them to be applied in more fields beyond high-temperature superconductivity. This work will inspire further experimental exploration and take an important step in the exploration of low-pressure stable high-temperature superconductors.